# Non-uniform electro-osmotic flow drives fluid–structure instability

Evgeniy Boyko[1§], Ran Eshel[1§], Amir D. Gat[1*] and Moran Bercovici[1,2*]

§Equal contribution

[1]Faculty of Mechanical Engineering, Technion - Israel Institute of Technology, Haifa, 3200003 Israel

[2]Department of Mechanical Engineering, The University of Texas at Austin, Austin, Texas 78712, USA

*corresponding authors: A.D.G.  (*amirgat@technion.ac.il*) and M.B. (*mberco@technion.ac.il*)

We demonstrate the existence of a fluid–structure instability arising from the interaction of electro-osmotic flow with an elastic substrate. Considering the case of flow within a soft fluidic chamber, we show that above a certain electric field threshold, negative gauge pressure induced by electro-osmotic flow causes the collapse of its elastic walls. We combine experiments and theoretical analysis to elucidate the underlying mechanism for instability and identify several distinct dynamic regimes. The understanding of this instability is important for the design of electrokinetic systems containing soft elements.

*Introduction.* Electro-osmotic flow (EOF) arises over electrically charged surfaces due to interaction of an externally applied electric field with the net charge in the electric double layer on a surface. Since its discovery by Reuss in 1809 [1], EOF has become a common method to manipulate fluids in microfluidic and lab-on-a-chip devices [2]. In many microfluidic applications, EOF acts against hydraulic resistance, resulting in an internal pressure distribution, which can be (gauge) positive or negative depending on the direction of the flow and the associated boundary conditions [2]. Since electro-osmotic flow rate scales as $\tilde{h}\tilde{u}_{EOF}$, and pressure driven flow rate scales as $\tilde{p}\tilde{h}^3/\tilde{\mu}\tilde{l}$, conservation of mass dictates a characteristic pressure of [3],

$$\tilde{p} \sim \frac{\tilde{\mu}\tilde{l}}{\tilde{h}^2}\tilde{u}_{EOF}, \tag{1}$$

where $\tilde{u}_{EOF}$ is the electro-osmotic slip velocity, $\tilde{h}$ is the height of the channel, $\tilde{l}$ is the characteristic stream-wise length-scale, and $\tilde{\mu}$ is the fluid viscosity.

Many microfluidic configurations are fabricated from soft materials such as poly(dimethylsiloxane) (PDMS) [4], and thus may deform due to fluid flow within the device. Such deformations were also recently studied in the context of EOF-driven flows by Mukherjee *et al.* [5], de Rutte *et al.* [6] and Boyko *et al.* [7]. In all of those studies, the fluid–structure interaction exhibited stable behavior.

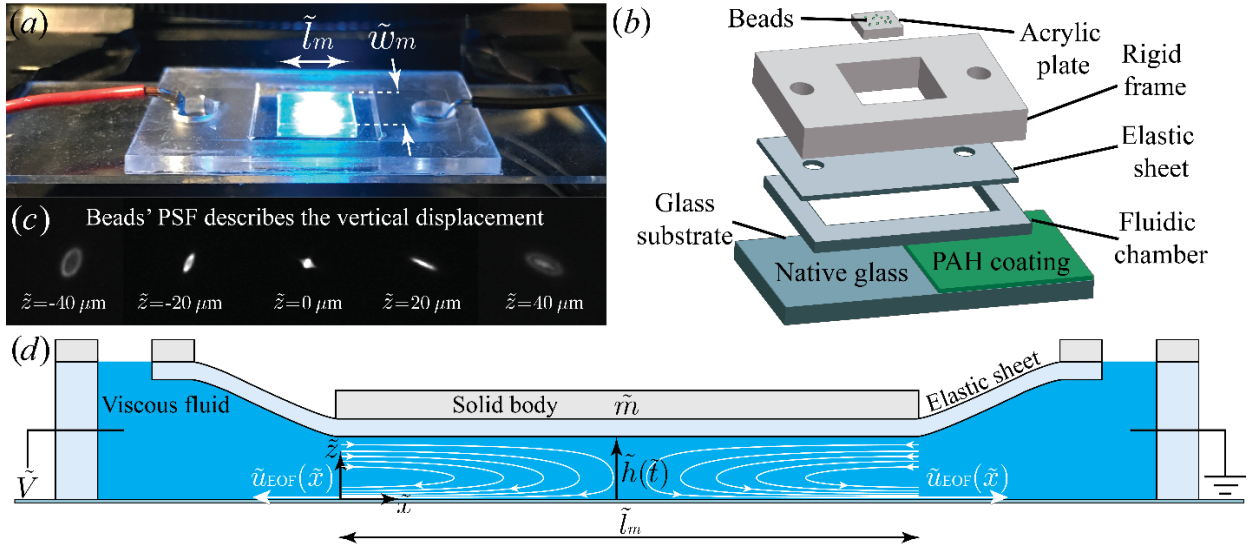

FIG. 1. *Illustration of the configuration used for experiments and modeling. (a) Image of the experimental device, (b) exploded isometric view showing its different layers, and (d) two-dimensional model and key parameters used for the theoretical analysis. The device consists of a microfluidic chamber chemically functionalized to produce non-uniform EOF. The chamber's floor is fixed while its ceiling can move vertically, supported by an elastic sheet. Upon application of an electric field, flow is driven from the center to the edges of the chamber, resulting in negative*

*pressure and downward motion of the ceiling. (c) We monitor the height of the ceiling, $\tilde{h}(\tilde{t})$, in time by capturing the change in the point spread function (PSF) of beads deposited on top of it, and observe its collapse onto the bottom surface when instability is triggered.*

In this Letter, we show for the first time that such electro-osmotic flow systems exhibit fluid–structure instability. Negative pressures induced by EOF lead to deformation of the elastic walls, decreasing the fluidic film thickness, $\tilde{h}$. In accordance with Eq. (1), this results in the pressure becoming increasingly negative, acting to further reduce the gap. Using experimental observations and theoretical predictions, we show the existence of an instability wherein small changes in the electric field lead to large changes in the deformation and ultimately the collapse of the elastic wall. We provide insight into the physical mechanisms underlying it and demonstrate that above a certain electric field threshold, the system switches from a stable behavior to an unstable one, characterized by a meta-stable bottleneck period. When the electric field is further increased, the bottleneck disappears and the system transitions to exhibit an immediate collapse of the elastic wall.

*Experimental.* To observe the dynamics of an elastic boundary subjected to negative pressure induced by non-uniform EOF, we designed an experimental system, shown in Fig. 1. The system consists of a $\tilde{h}_i = 171$ μm deep fluidic chamber with thin, 40 μm, elastic ceiling made of polydimethylsiloxane (PDMS). At the far edges of the 30 mm long chamber, there are two reservoirs through which the driving electric field is applied to the system. The bottom of the fluidic chamber is a glass slide, half of which is coated with poly(allylamine hydrochloride) (PAH) [8], a positively charged polyelectrolyte, changing the glass' native negative surface charge to a positive one. The elastic ceiling is supported everywhere by a rigid acrylic frame, except for a 15x15 mm region whose center is aligned with the surface charge discontinuity on the glass. On top of this region, we placed a 0.2 gr, 10x10 mm rigid acrylic plate, which at rest brings the liquid thickness to $\tilde{h}_0 = 94$ μm and stretches the elastic ceiling. We measure the vertical translation of the plate in time, $\tilde{h}(\tilde{t})$, by monitoring the change in the point spread function (PSF) of fluorescent micro-beads deposited on its top surface, as shown Fig 1(b). To enable measurement over the entire range of motion of the plate, we modify the PSF using a cylindrical lens, placed in front of the camera sensor [9]. At each time point, the image of each bead's PSF is compared against a pre-established calibration curve yielding its vertical position. We fill the fluidic chamber with 10 mM histidine providing simultaneously low conductivity and high buffering capacity, and applied a set of fixed voltages between the two reservoirs. The direction of the electric field is set such that the established EOF velocity is directed from the center of the chamber outward, thus inducing a negative pressure within the chamber.

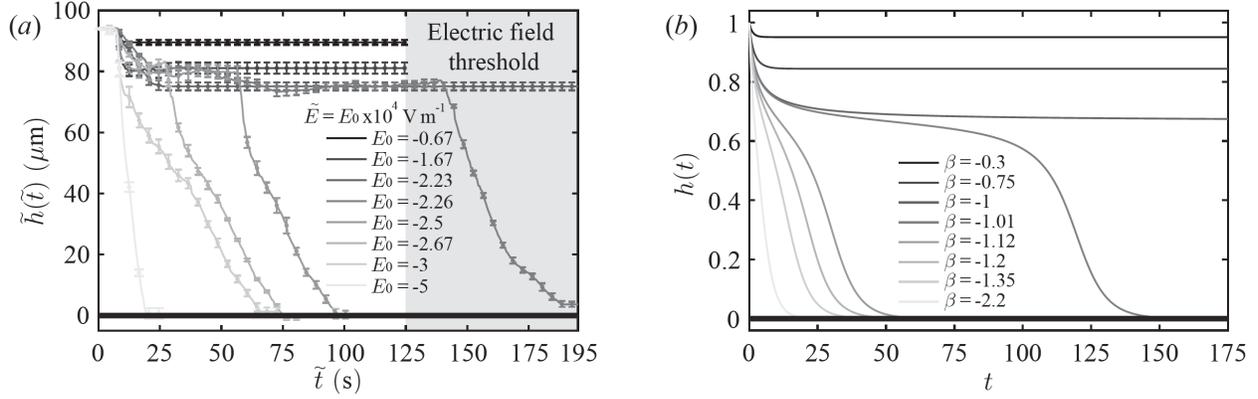

FIG. 2. *Time evolution of the fluidic gap resulting from non-uniform electro-osmotic flow. (a) The experimentally observed temporal evolution of the film thickness $\tilde{h}(\tilde{t})$ for various values of $\tilde{E}$ and (b) the corresponding theoretically predicted evolution of $h(t)$ for various values of $\beta$, indicating three distinct dynamic behaviors. Below the threshold values of $\tilde{E}_{CR} = -2.23 \times 10^4$ V m$^{-1}$ or $\beta_{CR} = -1$, the ceiling approaches a stable steady-state height. Setting $\tilde{E}$ or $\beta$ to slightly above their threshold values ($\tilde{E} = -2.26 \times 10^4$ V m$^{-1}$ and $\beta = -1.01$), triggers instability characterized by a bottleneck period. For higher values of $\tilde{E}$ or $\beta$ (e.g. for $\tilde{E} = -5 \times 10^4$ V m$^{-1}$ or $\beta = -2.2$), the bottleneck phase disappears and the ceiling immediately collapses onto the rigid floor. Error bars in (a) indicate a 95% confidence of the mean position of nine beads measured at each time point. The grayed-out region highlights two experiments in which the driving field differs by only 1.3%, capturing the threshold of instability and leading to a longer bottleneck due to the proximity to this value.*

*Experimental observations.* Figure 2(a) presents one data set of experimental measurements, showing the height of the plate relative to the bottom glass surface as a function of time for several applied voltages. Additional data sets exhibiting identical behavior are provided in the Supplementary Material [10]. We observe three distinct regimes for the dynamics of the plate; (I) Below a certain electric field threshold value (here approximately $\tilde{E}_{CR} = -2.23 \times 10^4$ V m$^{-1}$), the plate is pulled downward and achieves a new steady-state position where the hydrodynamic forces are balanced by the restoring force of the elastic membrane. (II) For electric fields above this threshold, the plate appears to be reaching a steady state as it lingers at a nearly fixed height for a significant duration of time, but finally, without any external interference to the system, it accelerates down and rapidly collapses onto the bottom surface. (III) As the electric field is increased this metastable 'bottleneck time', in which the plate descent is slowed down, shortens. For sufficiently high electric fields it completely disappears and the plate collapses to the floor immediately after application of the electric field. Figure 3(a) presents the height of the plate at the end of the experiment (after 195 s) as a function of the applied electric field, distinctly indicating the onset of instability. While for an electric field of $\tilde{E}_{CR} = -2.23 \times 10^4$ V m$^{-1}$ the system reaches a steady state at a moderate deformation, an increase of only 1.3% in the magnitude of the field results in an abrupt change in behavior, collapsing the plate onto the bottom surface. This can also be observed in the grayed-out region of Fig 2(a), where we allowed a longer measurement time in order to capture as accurately as possible the onset of instability.

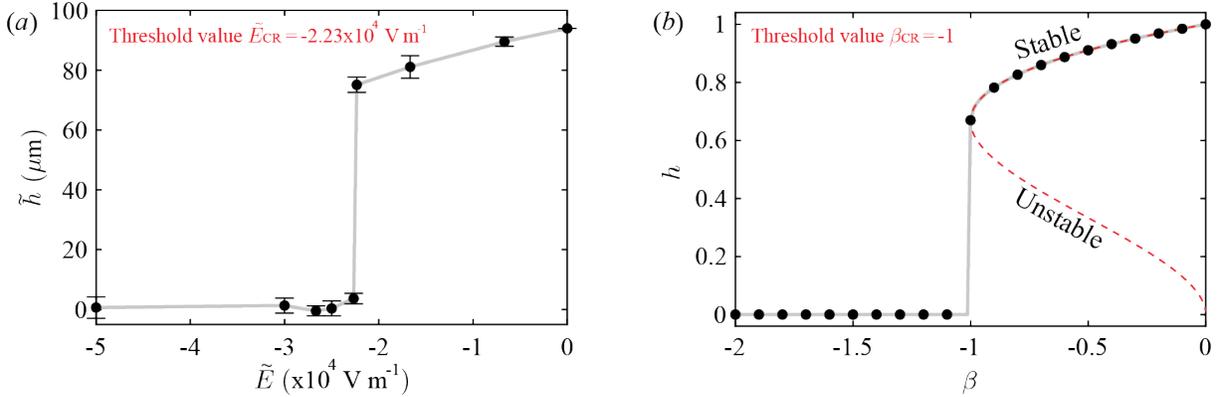

FIG. 3. *Theoretical and experimental observation of a threshold electric field for instability. (a) The experimentally observed height $\tilde{h}$ at the end of each experiment, as a function of applied electric field $\tilde{E}$. Error bars indicate a 95% confidence on the mean based on four experiments* [10]. *(b) The theoretically predicted steady-state height $h$ as a function of $\beta$. The red dashed line represents the results of a linear stability analysis, whereas black dots represent the results of the dynamic simulation. At $\beta_{CR} = -1$ the stable equilibrium intersects an unstable solution and disappears at a saddle-node bifurcation* [10]. *Above the threshold values of $\tilde{E}_{CR} = -2.23\times 10^4$ V m$^{-1}$ and $\beta_{CR} = -1$, no equilibrium solutions exist and the system exhibits instability which collapses the ceiling onto the floor. Gray lines in both sub-figures were added to guide the eye.*

*Theoretical model.* To provide further insight into the physical behavior of the system, we formulate a one-dimensional model describing the temporal evolution of the film thickness $\tilde{h}(\tilde{t})$. We consider a viscous liquid film confined between a rigid surface and a rigid body of mass $\tilde{m}$ and length $\tilde{l}_m$, placed on the top of an elastic sheet, which we model as a linear spring with stiffness $\tilde{k}$, as shown in Fig. 1(d). The liquid film is connected to two fluidic reservoirs at a distance $\tilde{L}$ from one another, through which the electric field is applied. We employ a Cartesian coordinate system ($\tilde{x}, \tilde{z}$), as indicated if Fig. 1(d). As in our experimental set-up, we prescribe a non-uniform electro-osmotic slip velocity $\tilde{u}_{EOF}(\tilde{x})$ on the bottom rigid surface, which in the thin-double-layer limit, can be described by the Helmholtz–Smoluchowski equation [11],

$$\tilde{u}_{EOF} = -\frac{\tilde{\varepsilon}\tilde{\zeta}\tilde{E}}{\tilde{\mu}}, \tag{2}$$

where $\tilde{\varepsilon}$ is the liquid permittivity, $\tilde{\zeta}(\tilde{x})$ is the zeta potential distribution on the surface, and $\tilde{E}$ is the imposed electric field. It is important to note that while the experimental system is inherently three dimensional, since the EOF-induced pressure gradients are primarily along the $\tilde{x}$-axis, we expect to capture the key properties of the system from a two-dimensional analysis in the $\tilde{x} - \tilde{z}$ plane.

Applying the lubrication approximation to the flow field, we relate the fluidic pressure $\tilde{p}$ to the gap height $\tilde{h}(\tilde{t})$, [10]

$$\frac{d\tilde{h}(\tilde{t})}{d\tilde{t}} - \frac{\tilde{h}(\tilde{t})^3}{12\tilde{\mu}} \frac{\partial^2 \tilde{p}(\tilde{x},\tilde{z},\tilde{t})}{\partial \tilde{x}^2} + \frac{1}{2}\tilde{h}(\tilde{t}) \frac{d\tilde{u}_{EOF}(\tilde{x})}{d\tilde{x}} = 0. \qquad (3)$$

Solving Eq. (3) for the pressure $\tilde{p}$ at $\tilde{z} = \tilde{h}$, and then integrating the result with respect to $\tilde{x}$ from 0 to $\tilde{l}_m$ while using the global mass conservation, provides the force $\tilde{F}_f$ that the fluid exerts on the rigid body (see [10]),

$$\tilde{F}_f(\tilde{t}) = -\frac{\tilde{\mu}\tilde{l}_m^3}{\tilde{h}(\tilde{t})^3}\frac{d\tilde{h}(\tilde{t})}{d\tilde{t}} + \frac{6\tilde{\mu}\tilde{B}}{\tilde{h}(\tilde{t})^2} + \chi\tilde{\rho}\tilde{g}\tilde{l}_m(\tilde{h}_i - \tilde{h}(\tilde{t})), \qquad (4)$$

where $\chi = 1 + [\tilde{l}_m / (\tilde{L} - \tilde{l}_m)]$ and $\tilde{B}$ is defined as

$$\tilde{B} = -\frac{\tilde{\varepsilon}\tilde{E}}{\tilde{\mu}}\left[\int_0^{\tilde{l}_m}\left[\int_0^{\tilde{x}}\zeta(\tilde{\xi})d\tilde{\xi}\right]d\tilde{x} - \frac{\tilde{l}_m}{2}\int_0^{\tilde{l}_m}\zeta(\tilde{\xi})d\tilde{\xi}\right]. \qquad (5)$$

The first term on the right-hand side of Eq. (4) represents the viscous resistance, the second term represents the electro-osmotic force which can be either attractive or repulsive depending on the sign of $\tilde{B}$, and the last term represents the restoring effect of the hydrostatic pressure.

Neglecting the plate's inertia, the force balance on the rigid plate, accounting for the fluidic force, as well as the elastic spring, yields a governing equation for the gap $\tilde{h}(\tilde{t})$, [10]

$$-\frac{\tilde{\mu}\tilde{l}_m^3}{\tilde{h}(\tilde{t})^3}\frac{d\tilde{h}(\tilde{t})}{d\tilde{t}} + \frac{6\tilde{\mu}\tilde{B}}{\tilde{h}(\tilde{t})^2} + \tilde{k}_g(\tilde{h}_0 - \tilde{h}(\tilde{t})) = 0, \qquad (6)$$

where $\tilde{k}_g = \tilde{k} + \chi\tilde{\rho}\tilde{g}\tilde{l}_m$ is the generalized spring stiffness consisting of contributions of elasticity and gravity, and $\tilde{h}_0 = \tilde{h}_i - (\tilde{m}\tilde{g} / \tilde{k}_g)$ is the liquid film thickness when the plate is at rest.

Equation (6) belongs to a family of non-linear evolution equations encountered in a range of instability problems such as electrostatic MEMS actuators [12], and elasto-capillarity coalescence [13]. Here the different terms in the Eq. (6) represent the coupling between viscous resistance, the electro-osmotic force and the restoring effects of the elasticity and gravity. As we show in the Supplementary Material [10], the case of constant current results in yet another variant of the equation, with different scaling for the actuation force ($\tilde{h}^{-3}$).

We introduce the non-dimensional variables $x = \tilde{x}/\tilde{l}_m$, $h = \tilde{h}/\tilde{h}_0$, and $t = \tilde{t}/\tilde{t}^*$, where $\tilde{t}^* = \tilde{\mu}\tilde{l}_m^3 / \tilde{k}_g \tilde{h}_0^3$ is the viscous–elastic time scale obtained by balancing the first and the third terms in Eq. (6). With this nondimensionalization, the governing equation (6) in non-dimensional form reads

$$\frac{1}{h^3}\frac{dh}{dt} = \frac{4}{27}\frac{\beta}{h^2} + 1 - h, \tag{7}$$

subject to the initial condition $h(t=0)=1$, where $\beta = 81\tilde{B}/(2\tilde{k}_g \tilde{h}_0^3 / \tilde{\mu})$ is the key governing non-dimensional parameter representing the ratio of electro-osmotic to elastic forces.

To understand the physical mechanisms underlying the instability, we perform a linear stability analysis of Eq. (7). We consider a small perturbation of the rigid body from its equilibrium height $h_{ss}$, satisfying $(4/27)\beta/h_{ss}^2 + 1 - h_{ss} = 0$, by setting $h(t) = h_{ss}(1 + \epsilon_0 e^{\sigma t})$, where $\epsilon_0 \ll 1$ is some small perturbation and $\sigma$ is the non-dimensional growth rate. We obtain that the growth rate $\sigma$ is [10]

$$\sigma = -\frac{8}{27}\beta - h_{ss}^3 = h_{ss}^2(2 - 3h_{ss}), \tag{8}$$

indicating that positive values of $\beta$, corresponding to positive (upward) deformation always result in a stable system, whereas negative values of $\beta$ may destabilize it. The critical steady-state height is thus $h_{ssCR} = 2/3$, corresponding to a threshold value of $\beta_{CR} = -1$, below which the system is unstable.

Figure 3(b) presents the steady-state height $h_{ss}$ as a function of $\beta$. The dashed red line represents the steady-state solutions of Eq. (7). For $\beta > \beta_{CR} = -1$, there are two real steady-state solutions, one of which is linearly stable and the other is linearly unstable. At $\beta_{CR} = -1$, these two solutions coincide and disappear at a saddle-node (fold) bifurcation with $h_{ssCR} = 2/3$. Black dots represent the results of a numerical simulation showing the collapse dynamics, consistent with the results of linear stability analysis.

To explain the experimentally observed temporal evolution of the film thickness showing strong dependence on the applied electric field (see Fig. 2(a)), we solve numerically the nonlinear evolution equation (7) for fixed values of $\beta$. It is convenient to discuss a normalized electric field difference, $\delta$, defined as

$$\delta = \frac{\tilde{E} - \tilde{E}_{CR}}{\tilde{E}_{CR}} = \frac{\beta - \beta_{CR}}{\beta_{CR}}, \tag{9}$$

so that $\beta = \beta_{CR}(1+\delta)$ and $\tilde{E} = \tilde{E}_{CR}(1+\delta)$, and the instability occurs for $\delta > 0$.

Figure 2(b) presents the time evolution of the gap $h(t)$ for different values of $\beta$. Our theoretical analysis identifies the three distinct dynamic behaviors observed in the experiments, with $\beta_{CR}$ ($\delta = 0$) serving as the threshold parameter. The analysis shows that a bottleneck behavior is obtained only for very small values of $\beta$ above this threshold, i.e. $0 < \delta \ll 1$. Furthermore, our analysis also explains the significant slow-down of the plate as it approaches the surface: the viscous resistance increases as $h^{-3}$, while the electro-osmotic attraction increases as $h^{-2}$. An asymptotic analysis for long times shows that this results in an exponential descent rate, for any $\delta > 0$ [10].

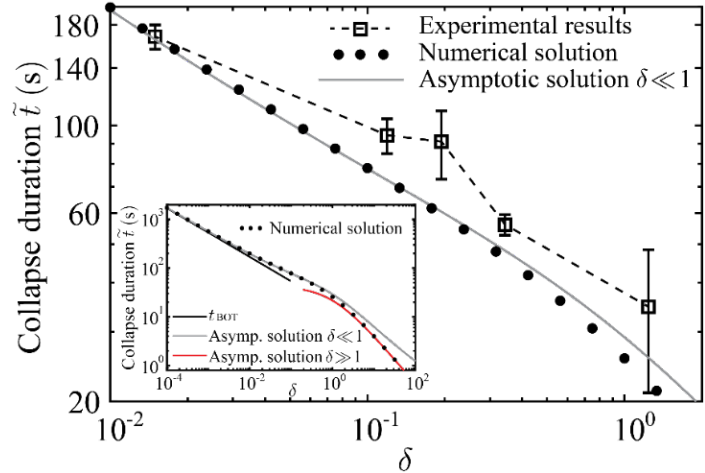

FIG. 4. *Collapse duration as a function of the normalized electric field difference $\delta$, corresponding to the time required for the system to reach 1% of its initial height after the onset of instability. Black squares and dots represent experimental and numerical results, respectively. Error bars indicate a 95% confidence on the mean based on four experiments [10]. To present the non-dimensional theoretical results in dimensional form we scale them by $\tilde{t}^* = 1.3866$ s, obtained by fitting the model to a single experimental data point at $\delta = 1.49 \times 10^{-2}$, $\tilde{t} = 168.2$ s. The gray line represents the asymptotic solution for $\delta \ll 1$. Inset: Theoretically predicted collapse duration for a wide range of $\delta$ values. Black dots represents the numerical results. The black line represents the prediction based on the bottleneck time alone in the limit $\delta \ll 1$, and the gray line represents the solution accounting also for the late time asymptotic behavior. The red line provides the asymptotic prediction for $\delta \gg 1$.*

As shown Fig. 2 both the theory and experiments indicate that the total collapse duration varies over at least one order of magnitude, and is strongly dependent on $\delta$. We here seek to quantify this collapse time and define it as the total time required for the plate to reach 1% of its initial height. Figure 4 compares the experimentally (squares) and theoretically predicted (dots and gray line) collapse duration as a function of $\delta$.

While our theoretical model does not account for various effects, such as three-dimensionality and nonlinear elastic effects, the model captures well the monotonic decrease of the collapse duration observed in the experiments. Figure 4 (inset) presents the theoretically predicted collapse duration for a wide range of $\delta$ values. For very small values of $\delta$, the total collapse time is dominated by the bottleneck time (i.e. the initial and final transients are very short in comparison), which is presented as the black line, denoted by $t_{BOT}$ and scaling as $\delta^{-1/2}$ [10]. The red curve provides the asymptotic solution for $\delta \gg 1$, where the collapse is purely exponential, which scales as $\delta^{-1}$ [10]. These time scales are similar to the ones obtained in overdamped MEMS devices [14,15]. Significant improvement in the analytically predicted collapse time can be obtained by accounting of the late time asymptotic behavior in the limit $\delta \ll 1$, provided by the gray line. Surprisingly, the approximation captures well the collapse time even for large values of delta, and exhibits a change in slope consistent with the numerical results.

*Summary and conclusions.* EOF systems are typically considered to be symmetric, in the sense that inversion of the electric field also inverts the direction of the flow. The coupling with elasticity breaks this symmetry, as is the case in other problems involving flow and elasticity, such as microswimmers and physiological flows (e.g. flow in the alveoli). This asymmetry is greatly enhanced by the instability we presented here, and the system can be considered as a deformation-based diode; for an electric field acting in a direction creating positive pressure and positive deformation, flow will be allowed through the channel. However, for an electric field acting in the opposite direction, the instability will be triggered, and the collapse of the channel walls will stop the flow. This instability could potentially also be leveraged as an actuation mechanism providing relatively large deformations and fast response, driven by an electric field, for a range of applications (e.g. soft robotics, configurable microfluidics, and adaptive optics).

While we studied here a single chamber, one could consider expansion of this study to multiple chambers, in parallel or in series, which interact with one another through both the fluid flow and the electric current distribution. Such a configuration would exhibit spatiotemporal propagation of the flow and electric fields, and of the resulting instability. One limit of such a distributed system would be an elastic substrate that can deform arbitrarily under a general surface potential distribution. The interplay between EOF-induced pressures and elastic deformation in such a case may lead to additional rich physics such as moving contact points involving triple phase contact lines.

**References**

[1] F. F. Reuss, Mem. Soc. Imp. Natur. Moscou **2**, 327 (1809).

[2] H. A. Stone, A. D. Stroock, and A. Ajdari, Annu. Rev. Fluid Mech. **36**, 381 (2004).

[3] D. J. Laser and J. G. Santiago, J. Micromech. Microeng. **14**, R35 (2004).

[4] G. M. Whitesides, Nature **442**, 368 (2006).

[5] U. Mukherjee, J. Chakraborty, and S. Chakraborty, Soft Matter **9**, 1562 (2013).

[6] J. M. de Rutte, K. G. Janssen, N. R. Tas, J. C. Eijkel, and S. Pennathur, Microfluid. Nanofluid. **20**, 150 (2016).

[7] E. Boyko, R. Eshel, K. Gommed, A. D. Gat, and M. Bercovici, J. Fluid Mech. **862**, 732 (2019).

[8] F. Paratore, E. Boyko, G. V. Kaigala, and M. Bercovici, Phys. Rev. Lett. **122**, 224502 (2019).

[9] R. Barnkob, C. J. Kähler, and M. Rossi, Lab Chip **15**, 3556 (2015).

[10] See Supplemental Material at [XXX] for further details of experiments and theoretical calculations, which includes Refs. [22–25]., (n.d.).

[11] R. J. Hunter, *Foundations of Colloid Science* (Oxford University Press, New York, 2001).

[12] G. M. Rebeiz, *RF MEMS: Theory, Design, and Technology* (John Wiley & Sons, 2004).

[13] K. Singh, J. R. Lister, and D. Vella, J. Fluid Mech. **745**, 621 (2014).

[14] R. K. Gupta and S. D. Senturia, in *Proc. IEEE Int. Workshop on MEMS* (IEEE, 1997), pp. 290–294.

[15] M. Gomez, D. E. Moulton, and D. Vella, J. Micromech. Microeng. **28**, 015006 (2017).
Page **10** of **10****References**

[1] F. F. Reuss, Mem. Soc. Imp. Natur. Moscou **2**, 327 (1809).

[2] H. A. Stone, A. D. Stroock, and A. Ajdari, Annu. Rev. Fluid Mech. **36**, 381 (2004).

[3] D. J. Laser and J. G. Santiago, J. Micromech. Microeng. **14**, R35 (2004).

[4] G. M. Whitesides, Nature **442**, 368 (2006).

[5] U. Mukherjee, J. Chakraborty, and S. Chakraborty, Soft Matter **9**, 1562 (2013).

[6] J. M. de Rutte, K. G. Janssen, N. R. Tas, J. C. Eijkel, and S. Pennathur, Microfluid. Nanofluid. **20**, 150 (2016).

[7] E. Boyko, R. Eshel, K. Gommed, A. D. Gat, and M. Bercovici, J. Fluid Mech. **862**, 732 (2019).

[8] F. Paratore, E. Boyko, G. V. Kaigala, and M. Bercovici, Phys. Rev. Lett. **122**, 224502 (2019).

[9] R. Barnkob, C. J. Kähler, and M. Rossi, Lab Chip **15**, 3556 (2015).

[10] See Supplemental Material at [XXX] for further details of experiments and theoretical calculations, which includes Refs. [22–25]., (n.d.).

[11] R. J. Hunter, *Foundations of Colloid Science* (Oxford University Press, New York, 2001).

[12] G. M. Rebeiz, *RF MEMS: Theory, Design, and Technology* (John Wiley & Sons, 2004).

[13] K. Singh, J. R. Lister, and D. Vella, J. Fluid Mech. **745**, 621 (2014).

[14] R. K. Gupta and S. D. Senturia, in *Proc. IEEE Int. Workshop on MEMS* (IEEE, 1997), pp. 290–294.

[15] M. Gomez, D. E. Moulton, and D. Vella, J. Micromech. Microeng. **28**, 015006 (2017).